\documentclass[draftclsnofoot,onecolumn,12pt]{IEEEtran}

\usepackage[T1]{fontenc}% optional T1 font encoding

\ifCLASSINFOpdf
\else
\fi
\usepackage{amsmath}
\usepackage{amsfonts}
\usepackage{mathtools}
\usepackage{color}
\usepackage[cmintegrals]{newtxmath}
\begin{document}
\title{Performance Analysis of Massive MIMO with Low-Resolution ADCs}

\author{Chao Wei and Zaichen Zhang, \emph{Senior Member, IEEE}% <-this % stops a space
\thanks{The authors are with the National Mobile Communications Research Laboratory, Southeast University, Nanjing 210096, China (emails: weichao@seu.edu.cn; zczhang@seu.edu.cn).}}

\maketitle

\begin{abstract}
%In this letter, we consider the uplink performance of massive multiple-input-multiple-output (MIMO) systems, where the base stations (BS) employ low-resolution analog-to-digital converters (ADCs) and minimum mean-square error (MMSE) receiver. By modeling the quantization noise of low-resolution ADCs as an additive quantization noise, the high performance MMSE receiver can be carefully designed to take both additive white Gaussian noise (AWGN) and quantization noise into account, which can work well for both cases of uniform resolution ADCs and non-uniform resolution ADCs. Then, based on random matrix theory and the proposed MMSE receiver, we derive the asymptotic equivalent expression of the uplink spectral efficiency (SE) of massive MIMO systems. Numerical results show the tightness of the asymptotic equivalent expression of the uplink SE, and the low-resolution ADCs can still achieve the satisfying uplink SE by increasing the number of antennas at BS.
The uplink performance of massive multiple-input-multiple-output (MIMO) systems where the base stations (BS) employ low-resolution analog-to-digital converters (ADCs) is analyzed.
A high performance MMSE receiver 
that takes both additive white Gaussian noise (AWGN) and quantization noise into consideration is designed, which works well
for both cases of uniform resolution ADCs and non-uniform resolution ADCs.
With the proposed MMSE receiver, we then employ the random matrix theory
to derive the 
asymptotic equivalent of the uplink spectral efficiency (SE) of the system.
Numerical results show the tightness of the asymptotic equivalent of the uplink SE, and massive MIMO with low-resolution ADCs can still achieve the satisfying uplink SE by increasing the number of antennas at BS.

\end{abstract}


% \fi
\IEEEpeerreviewmaketitle

\section{Introduction}
\IEEEPARstart{M}{assive} 
multiple-input multiple-output (MIMO) has been considered as one of the key enabling technologies for 5G wireless communications systems, and thus has recently drawn lots of research interest. %Significant spectral efficiency (SE) and energy efficiency (EE) gains can be achieved when the base station (BS) is equipped with large number of antennas \cite{Andrews}, \cite{Ngo}. 
It could bring significant spectral efficiency (SE) and energy efficiency (EE) gains when the base station (BS) has a large number of antennas \cite{Andrews}, \cite{Ngo}.
%With the aid of 
Due to the large number of antennas at BS, a simple linear receiver can be employed to achieve a high SE, and the impact of imperfect hardware can also be mitigated. Most of the current works concerning massive MIMO systems assume each antenna at BS is equipped with a high-resolution ADC. 
However, high-resolution ADCs can incur substantial power assumption and high-cost hardware implementation \cite{Le}, which may be not affordable
%is not feasible 
in practical massive MIMO systems.
%This fact motives us to study low-resolution ADCs that hold the favorable property of low cost, low power consumption and feasibility of implementation \cite{Le}.
This motives us to study the use of low-resolution ADCs to understand their impact on the performance of the system.

Several recent works have investigated the effects of low-resolution ADCs on the performance of massive MIMO systems \cite{Bai}-\cite{Zhang2016}. 
%In these works, the common widely used method to describe the effect of low-resolution ADCs, is the additive quantization noise model (AQNM) \cite{Bai}.
In these works, the effect of low-resolution ADCs is described by using the additive quantization noise model (AQNM) \cite{Bai}.
Using AQNM, 
%\cite{Fan} 
Fan {\it et al.} \cite{Fan}
derive an approximate analytical expression for the uplink achievable rate of massive MIMO systems over Rayleigh fading channels when low-resolution ADCs and the maximal-ratio combining (MRC) technique are used at the receiver; 
%\cite{Zhang2016} 
Zhang {\it et al.} \cite{Zhang2016}
derive tractable and exact approximation expressions for the uplink SE of massive MIMO systems with low-resolution ADCs and MRC over Rician fading channels, where both perfect and imperfect channel state information (CSI) are considered. 
%However, \cite{Fan} and \cite{Zhang2016} only considered uniform resolution ADCs and MRC receiver, \cite{Zhang2016}
However, only uniform resolution ADCs and MRC receiver are considered in \cite{Fan} and \cite{Zhang2016}, which cannot be applied to  analyze  the  uplink  SE  performance  of  massive  MIMO systems  with  non-uniform  resolution  ADCs  and  MMSE  receiver.
In \cite{Liang}, a mixed-ADC architecture for massive MIMO systems is proposed, in which the BS antennas are equipped with both high-resolution ADCs and low-resolution ADCs. Under mixed-ADC massive MIMO with MRC receiver, the closed-form approximate expressions for the uplink achievable rate are derived for both perfect and imperfect CSI when the Rician fading channels are considered \cite{Zhang2017}. In \cite{Dong}, the asymptotic equivalent of the uplink SE is obtained when non-uniform resolution ADCs and MMSE receivers are used at the BS of massive MIMO systems. The MMSE receiver designed in \cite{Dong} could cause significant SE loss 
for 
%when 
massive MIMO systems with non-uniform resolution ADCs. 
%This is because \cite{Dong} 
This is because it does not consider non-uniform quantization noise incurred by non-uniform resolution ADCs.   

In this letter, with AQNM, we first design a proper MMSE receiver
%we carefully design the MMSE receiver 
for the uplink massive MIMO system with low-resolution ADCs. The proposed MMSE receiver takes  both additive white Gaussian noise (AWGN) and quantization noise into account. It can work well for both cases of uniform resolution ADCs and non-uniform resolution ADCs. % Then the tight asymptotic equivalent expression of the uplink SE is derived based on random matrix theory. At last numerical results verify our theoretical analysis.
We then derive a tight asymptotic equivalent of the uplink SE using random matrix theory. Finally, numerical results are obtained to verify our theoretical analysis. 

Notations: Boldface lower and upper-case symbols represent vectors and matrices, respectively. The ${N \times N}$ identity matrix reads as ${{\bf{I}}_N}$. let ${{\bf{X}}^H}$ and ${\mathop{\rm tr}\nolimits} \left( {\bf{X}} \right)$ be the conjugate transpose and trace of a matrix ${\bf{X}}$. ${\mathop{\rm diag}\nolimits} \left( {\bf{X}} \right)$ keeps only the diagonal entries of ${\bf{X}}$. A random vector ${\bf{x}} \sim {\cal C}{\cal N}\left( {{\bf{m}},{\bf{V}}} \right)$ is complex Gaussian distributed with mean vector ${\bf{m}}$ and covariance matrix ${\bf{V}}$. $\overset{a.s.}{\longrightarrow}$ denotes almost sure convergence.

\section{System Model}
Consider an uplink massive MIMO system with low-resolution ADCs, where the BS equipped with $M$ antennas receives signal from $K$ single-antenna users using the same time-frequency resources. The received signal ${\bf{y}} \in {{\mathbb{C}}^{M \times 1}} $ at BS can be written as
\begin{align} \label{EQ1}
{\bf{y}} = \sqrt {{p_u}} {\bf{Gx}} + {\bf{n}},
\end{align}
where ${p_u}$ is the average transmit power of each user, ${\bf{x}} \in {{\mathbb{C}}^{K \times 1}}$ is the transmit vector for all $K$ users, and ${\bf{n}} \sim {\cal C}{\cal N} \left( {0,{{\bf{I}}_M}} \right)$ is the normalized AWGN vector. The input covariance matrix ${{\bf{R}}_{\bf{x}}} = {{\bf{I}}_K}$ is assumed. The channel matrix 
${\bf{G}} \in {{\mathbb{C}}^{M \times K}}$ between the BS and all users is given as
\begin{align} \label{EQ2}
{\bf{G}} = {\bf{H}}{{\bf{D}}^{{1 \mathord{\left/
 {\vphantom {1 2}} \right.
 \kern-\nulldelimiterspace} 2}}},
\end{align}
where $\bf{H}$ denotes the small-scale fading channel matrix and ${\bf{D}}$ is the $K \times K$ diagonal matrix with diagonal entries $\left\{ {{\beta _k}} \right\}$ to model the large-scale fading. Rayleigh fading is assumed here, thus without loss of generality, the elements of ${\bf{H}}$ can be considered to follow distribution ${\cal C}{\cal N} \left( {0,1} \right)$. 

As the quantization error can be well approximated as a linear gain with AQNM \cite{Bai}, the output vector of low-resolution ADCs can be expressed as
\begin{align}   \label{EQ3}
{{\bf{y}}_q} = {\bf{Q}}\left( {\bf{y}} \right) \approx \Lambda {\bf{y}} + {{\bf{n}}_q} = \sqrt {{p_u}} \Lambda {\bf{Gx}} + \Lambda {\bf{n}} + {{\bf{n}}_q},
\end{align}
where ${\bf{Q}}\left(  \cdot  \right)$ is the quantizer function which applies component-wise and separately to the real and imaginary parts, and ${{\bf{n}}_q}$ is the additive Gaussian quantization noise vector which is uncorrelated with ${\bf{y}}$. The linear gain matrix $\Lambda$ is an $M \times M$ diagonal matrix with entries $\left\{ {{\alpha _m}} \right\}$, and 
%we have
the linear gain coefficient ${\alpha _m} = 1 - {\rho _m}$, where ${\rho _m}$ is the inverse of the signal-to-quantization-noise ratio. Given a fixed channel realization ${\bf{G}}$, the covariance matrix of ${{\bf{n}}_q}$ can be written as \cite{Fan} 
\begin{align} \label{EQ4}
\begin{array}{l}
{{\bf{R}}_{{{\bf{n}}_q}}} = {\rm E}\left( {{{\bf{n}}_q}{\bf{n}}_q^H\left| {\bf{G}} \right.} \right)\\
{\rm{~~~~~= }}\Lambda \left( {{{\bf{I}}_M} - \Lambda } \right){\bf{diag}}\left( {{p_u}{\bf{G}}{{\bf{G}}^H} + {{\bf{I}}_M}} \right).
\end{array}
\end{align}

\section{Asymptotic Equivalent of the Uplink Spectral Efficiency} \label{Section 3}
In this section, using random matrix theory, we derive the asymptotic equivalent of the uplink SE for massive MIMO systems with low-resolution ADCs, when the MMSE receiver is considered. Here the perfect CSI is known at the BS. First, the proposed high-performance MMSE receiver is designed as
\begin{align}  \label{EQ5}
{\bf{W}} = {{\bf{G}}^H}{\Lambda ^H}{\left( {\Lambda {\bf{G}}{{\bf{G}}^H}{\Lambda ^H} + {\bf{Z}} + \theta {{\bf{I}}_M}} \right)^{ - 1}},
\end{align}
where ${\bf{Z}} = \frac{1}{{{p_u}}}{{\bf{\bar R}}_{{{\bf{n}}_q}}} = \frac{1}{{{p_u}}}\Lambda \left( {{{\bf{I}}_M} - \Lambda } \right)\left( {{p_u}{\bf{tr}}\left( {\bf{D}} \right) + 1} \right)$, and $\theta  = {1 \mathord{\left/
 {\vphantom {1 {{p_u}}}} \right.
 \kern-\nulldelimiterspace} {{p_u}}}$. ${{\bf{\bar R}}_{{{\bf{n}}_q}}}$ is the expectation of ${{\bf{R}}_{{{\bf{n}}_q}}}$ with respect to ${\bf{G}}$. Thus, we can see that the proposed MMSE receiver can suppress both AWGN and quantization noise at the same time. This proposed MMSE receiver will be used in the following analysis. By multiplying ${\bf{W}}$ with ${{\bf{y}}_q}$, the quantized output vector is processed as
\begin{align} \label{EQ6}
    {\bf{r}} = {\bf{W}}{{\bf{y}}_q} = \sqrt {{p_u}} {\bf{W}}\Lambda {\bf{Gx}} + {\bf{W}}\Lambda {\bf{n}} + {\bf{W}}{{\bf{n}}_q}.
\end{align}
The output signal for the $k{\rm{th}}$ user can be expressed as
\begin{align}  \label{EQ7}
    \begin{array}{l}
{{\bf{r}}_k} = \sqrt {{p_u}} {\bf{g}}_k^H{\Lambda ^H}{\bf{V}}\Lambda {{\bf{g}}_k}{x_k} + \sqrt {{p_u}} \sum\limits_{i \ne k} {{\bf{g}}_k^H{\Lambda ^H}{\bf{V}}\Lambda {{\bf{g}}_i}{x_i}} \\
{\rm{~~~~~~+  }}{\bf{g}}_k^H{\Lambda ^H}{\bf{V}}\Lambda {\bf{n}} + {\bf{g}}_k^H{\Lambda ^H}{\bf{V}}{{\bf{n}}_q},
\end{array}
\end{align}
where ${\bf{V}} = {\left( {\Lambda {\bf{G}}{{\bf{G}}^H}{\Lambda ^H} + {\bf{Z}} + \theta {{\bf{I}}_M}} \right)^{ - 1}}$, ${{\bf{g}}_i}$ is the $i{\bf{th}}$ column of ${\bf{G}}$ and ${x_i}$ is the $i{\bf{th}}$ element of ${\bf{x}}$. The achievable uplink rate of the $k{\rm{th}}$ user is thus expressed as
\begin{align}  \label{EQ8}
    {R_k} = {\log _2}\left( {1 + \frac{{{p_u}{{\left| {{\bf{g}}_k^H{\Lambda ^H}{\bf{V}}\Lambda {{\bf{g}}_k}} \right|}^2}}}{{{I_k}}}} \right),
\end{align}
where the variance of the interference-plus-noise term is 
\begin{align}  \label{EQ9}
\begin{array}{l}
{I_k} = {p_u}\sum\limits_{i \ne k} {{{\left| {{\bf{g}}_k^H{\Lambda ^H}{\bf{V}}\Lambda {{\bf{g}}_i}} \right|}^2}}  + {\bf{g}}_k^H{\Lambda ^H}{\bf{V}}\Lambda {\Lambda ^H}{\bf{V}}\Lambda {{\bf{g}}_k}\\
{\rm{~~~~~+ }}{\bf{g}}_k^H{\Lambda ^H}{\bf{V}}{{\bf{R}}_{{{\bf{n}}_q}}}{\bf{V}}\Lambda {{\bf{g}}_k}
\end{array}.
\end{align}
Now the sum uplink SE of the whole system is obtained as
\begin{align}   \label{EQ10}
{R_{sum}} = \sum\limits_k {{R_k}}.
\end{align}

To obtain the asymptotic equivalent of ${R_k}$, we first derive the asymptotic equivalent for the power of the signal of interest and the three parts of ${I_k}$. Then, the asymptotic equivalent of ${R_k}$ is given as follows:

\emph{Lemma 1:} When $M,K \to \infty $ with fixed ratio, the asymptotic equivalent of the uplink SE in Eq. (\ref{EQ8}) is written as
\begin{align}  \label{EQ11}
{\bar R_k} = \frac{{{p_u}{\beta _k}{{\left| {{\Gamma _1}} \right|}^2}}}{{{p_u}\sum\limits_{i \ne k} {{{\left| {\frac{1}{{1 + {{{\beta _i}{\Gamma _1}} \mathord{\left/
 {\vphantom {{{\beta _i}{\Gamma _1}} M}} \right.
 \kern-\nulldelimiterspace} M}}}} \right|}^2}{\beta _i}{\Gamma _2} + {\Gamma _2} + {\Gamma _3}} }},
\end{align}
where ${\Gamma _1}$, ${\Gamma _2}$ and ${\Gamma _3}$ are calculated as Eqs. (14), (18) and (23), respectively.

\emph{Proof:} For the signal of interest, using Sherman-Morrison formula, we can obtain
\begin{align}   \label{EQ12}
{\bf{g}}_k^H{\Lambda ^H}{\bf{V}}\Lambda {{\bf{g}}_k} = \frac{{{\bf{g}}_k^H{\Lambda ^H}{{\left( {\Lambda {{\bf{G}}_{ - k}}{\bf{G}}_{ - k}^H{\Lambda ^H} + {\bf{Z}} + \theta {{\bf{I}}_M}} \right)}^{ - 1}}\Lambda {{\bf{g}}_k}}}{{1 + {\bf{g}}_k^H{\Lambda ^H}{{\left( {\Lambda {{\bf{G}}_{ - k}}{\bf{G}}_{ - k}^H{\Lambda ^H} + {\bf{Z}} + \theta {{\bf{I}}_M}} \right)}^{ - 1}}\Lambda {{\bf{g}}_k}}},
\end{align}
where ${{\bf{G}}_{ - k}}$ denotes that the $k{\bf{th}}$ column of ${\bf{G}}$ is removed from ${\bf{G}}$. Through \emph{Theorem 1} in {\bf{Appendix A}}, we have
\begin{equation}
{\bf{g}}_k^H{\Lambda ^H}{\left( {\Lambda {{\bf{G}}_{ - k}}{\bf{G}}_{ - k}^H{\Lambda ^H} + {\bf{Z}} + \theta {{\bf{I}}_M}} \right)^{ - 1}}\Lambda {{\bf{g}}_k} 
\nonumber \\
\end{equation}
\begin{equation}
\overset{(a)}{\longrightarrow} {\bf{tr}}\left( {{\beta _k}\Lambda {\Lambda ^H}{{\bf{V}}_{\left( k \right)}}} \right) 
\nonumber \\
{\longrightarrow} {\beta _k}{\bf{tr}}\left( {\Lambda {\Lambda ^H}{\bf{V}}} \right)
\end{equation}
\begin{equation} \label{EQ13}
\overset{a.s.}{\longrightarrow} {\beta _k}{\bf{tr}}\left( {\Lambda {\Lambda ^H}{\bf{S}}\left( {\frac{\theta }{M}} \right)} \right) = {\beta _k}{\Gamma _1},
\end{equation}
where (a) follows from \emph{Lemma 4} in \cite{Wagner}, ${{\bf{V}}_{\left( k \right)}} = {\left( {\Lambda {{\bf{G}}_{ - k}}{\bf{G}}_{ - k}^H{\Lambda ^H} + {\bf{Z}} + \theta {{\bf{I}}_M}} \right)^{ - 1}}$ and
\begin{equation} \label{EQ14}
{\Gamma _1} = {\bf{tr}}\left( {\Lambda {\Lambda ^H}{\bf{S}}\left( {\frac{\theta }{M}} \right)} \right).
\end{equation}
With Eq. (\ref{EQ13}), the asymptotic equivalent for the power of the signal of interest can be given as
\begin{equation}   \label{EQ15}
{\left| {{\bf{g}}_k^H{\Lambda ^H}{{\bf{V}}^{ - 1}}\Lambda {{\bf{g}}_k}} \right|^2}
\overset{a.s.}{\longrightarrow} \left| {\frac{{{\beta _k}{\Gamma _1}}}{{1 + {\beta _k}{\Gamma _1}}}} \right|.
\end{equation}

For the inter-user interference part, by using Sherman-Morrison formula twice, we have
\begin{equation}
{\left| {{\bf{g}}_k^H{\Lambda ^H}{{\bf{V}}^{ - 1}}\Lambda {{\bf{g}}_i}} \right|^2} = {\left| {\frac{1}{{1 + {\bf{g}}_k^H{\Lambda ^H}{{\bf{V}}_{\left( k \right)}}\Lambda {{\bf{g}}_k}}}} \right|^2} \cdot {\left| {\frac{1}{{1 + {\bf{g}}_i^H{\Lambda ^H}{{\bf{V}}_{\left( {ki} \right)}}\Lambda {{\bf{g}}_i}}}} \right|^2}
\nonumber \\
\end{equation}
\begin{equation}  \label{EQ16}
{\rm{~~~~~~~~~}} \cdot {\bf{g}}_k^H{\Lambda ^H}{{\bf{V}}_{\left( {ki} \right)}}\Lambda {{\bf{g}}_i}{\bf{g}}_i^H{\Lambda ^H}{{\bf{V}}_{\left( {ki} \right)}}\Lambda {{\bf{g}}_k}.
\end{equation}
Through \emph{Theorem 2} in {\bf{Appendix A}}, we have
\begin{equation}
{\bf{g}}_k^H{\Lambda ^H}{{\bf{V}}_{\left( {ki} \right)}}\Lambda {{\bf{g}}_i}{\bf{g}}_i^H{\Lambda ^H}{{\bf{V}}_{\left( {ki} \right)}}\Lambda {{\bf{g}}_k}
\nonumber \\
\end{equation}
\begin{equation}
\overset{(a)}{\longrightarrow} {\bf{tr}}\left( {{\beta _k}{\beta _i}{\Lambda ^H}{{\bf{V}}_{\left( {ki} \right)}}\Lambda {\Lambda ^H}{{\bf{V}}_{\left( {ki} \right)}}\Lambda } \right) {\longrightarrow} {\beta _k}{\beta _i}{\bf{tr}}\left( {\Lambda {\Lambda ^H}{\bf{V}}\Lambda {\Lambda ^H}{\bf{V}}} \right)
\nonumber \\
\end{equation}
\begin{equation} \label{EQ17}
\overset{a.s.}{\longrightarrow} {\beta _k}{\beta _i}{\bf{tr}}\left( {\Lambda {\Lambda ^H}{\bf{S'}}\left( {\frac{\theta }{M}} \right)} \right) = {\beta _k}{\beta _i}{\Gamma _2}, 
\end{equation}
where
\begin{equation}  \label{EQ18}
{\Gamma _2} = {\bf{tr}} \left( {\Lambda {\Lambda ^H}{\bf{S'}}\left( {\frac{\theta }{M}} \right)} \right).
\end{equation} 
Thus as $M,K \to \infty $, the asymptotic equivalent  of the inter-user interference part is expressed as
\begin{equation} \label{EQ19}
{\left| {{\bf{g}}_k^H{\Lambda ^H}{{\bf{V}}^{ - 1}}\Lambda {{\bf{g}}_i}} \right|^2}\
\overset{a.s.}{\longrightarrow} {\left| {\frac{1}{{1 + {\beta _k}{\Gamma _1}}}} \right|^2}{\left| {\frac{1}{{1 + {\beta _i}{\Gamma _1}}}} \right|^2}{\beta _k}{\beta _i}{\Gamma _2}.
\end{equation}

For the Gaussian noise part, again by using Sherman-Morrison formula, we have 
\begin{align}   \label{EQ20}
{\bf{g}}_k^H{\Lambda ^H}{\bf{V}}\Lambda {\Lambda ^H}{\bf{V}}\Lambda {{\bf{g}}_i} = {~~~~~~~~~~~~~~~~~~~~~~~~~~~~~} \nonumber \\ 
{\rm{ }}{\left| {\frac{1}{{1 + {\bf{g}}_k^H{\Lambda ^H}{{\bf{V}}_{\left( k \right)}}\Lambda {{\bf{g}}_k}}}} \right|^2}{\bf{g}}_k^H{\Lambda ^H}{{\bf{V}}_{\left( k \right)}}\Lambda {\Lambda ^H}{{\bf{V}}_{\left( k \right)}}\Lambda {{\bf{g}}_k}.
\end{align}
Then, using Eqs. (\ref{EQ13}) and (\ref{EQ17}) , the asymptotic equivalent of the Gaussian noise part is expressed as 
\begin{equation}  \label{EQ21}
{\bf{g}}_k^H{\Lambda ^H}{\bf{V}}\Lambda {\Lambda ^H}{\bf{V}}\Lambda {{\bf{g}}_i} 
\overset{a.s.}{\longrightarrow}
{\left| {\frac{1}{{1 + {\beta _k}{\Gamma _1}}}} \right|^2}{\beta _k}{\Gamma _2}.
\end{equation}

The quantization noise part contains ${{\bf{R}}_{{{\bf{n}}_q}}}$ that is dependent on the channel matrix ${\bf{G}}$. We use ${{\bf{\bar R}}_{{{\bf{n}}_q}}}$ to replace ${{\bf{R}}_{{{\bf{n}}_q}}}$ in ${\bf{g}}_k^H{\Lambda ^H}{\bf{V}}{{\bf{R}}_{{{\bf{n}}_q}}}{\bf{V}}\Lambda {{\bf{g}}_k}$ and the effect of this replacement on the asymptotic equivalent is negligible. Similar to the derivation of the Gaussian noise part, the asymptotic equivalent of the quantization noise part is obtained as
\begin{align}  \label{EQ22}
{\bf{g}}_k^H{\Lambda ^H}{\bf{V}}{{\bf{R}}_{{{\bf{n}}_q}}}{\bf{V}}\Lambda {{\bf{g}}_k} \approx {~~~~~~~~~~~~~~~~~~~~~~~~~~~~~~~~~~~~~~~~~~~~} \nonumber \\
{\left| {\frac{1}{{1 + {\bf{g}}_k^H{\Lambda ^H}{{\bf{V}}_{\left( k \right)}}\Lambda {{\bf{g}}_k}}}} \right|^2}{\bf{g}}_k^H{\Lambda ^H}{{\bf{V}}_{\left( k \right)}}{{{\bf{\bar R}}}_{{{\bf{n}}_q}}}{{\bf{V}}_{\left( k \right)}}\Lambda {{\bf{g}}_k}  \nonumber \\
\overset{a.s.}{\longrightarrow} 
{\left| {\frac{1}{{1 + {\beta _k}{\Gamma _1}}}} \right|^2}{\beta _k}{\bf{tr}}\left( {{{{\bf{\bar R}}}_{{{\bf{n}}_q}}}{\bf{S'}}\left( {\frac{\theta }{M}} \right)} \right) = {\left| {\frac{1}{{1 + {\beta _k}{\Gamma _1}}}} \right|^2}{\beta _k}{\Gamma _3},
\end{align}
where 
\begin{equation}    \label{EQ23}
{\Gamma _3} = {\bf{tr}}\left( {{{{\bf{\bar R}}}_{{{\bf{n}}_q}}}{\bf{S'}}\left( {\frac{\theta }{M}} \right)} \right).
\end{equation}

Until now, substituting Eqs. (\ref{EQ15}), (\ref{EQ19}), (\ref{EQ21}) and (\ref{EQ22}) into Eq. (\ref{EQ8}) and after some mathematics manipulations, we can get the \emph{Lemma 1}. {~~~~~~~~~~~}$\blacksquare$

Note that \emph{Lemma 1} is a general result and an exact asymptotic equivalent of the uplink SE for both uniform resolution ADCs and non-uniform resolution ADCs in massive MIMO systems. In addition, our proposed MMSE receiver shows better performances than that in [9], which is verified in simulation. Furthermore, the closed-form asymptotic equivalent of the uplink SE only depends on the statistics parameters of the channel that keeps static within a coherence time block; therefore it can be readily used to analytically evaluate the effects of low-resolution ADCs, the number of antennas $M$, and the transmit power ${p_u}$ on the considered system's performance.

\section{Numerical results}
%In this section, we provide numerical results to verify \emph{Lemma 1} in Sec. \ref{Section 3}.  We consider a hexagonal cell with radius of 1000 meters. The users are assumed to be uniformly distributed over the cell, while the smallest distance between the users and the BS is ${r_{\min }} = 100$ meters. The pathloss is expressed as $r_n^{ - v}$, where ${r_n}$ is the distance from the $n{\bf{th}}$ user to the BS and $v = 3.8$ is the pathloss exponent. The shadowing is modeled as a log-normal random variable ${s_n}$
%with standard deviation of 8dB. Now, the large-scale fading can be given as ${\beta _n} = {s_n}{\left( {{{{r_n}} \mathord{\left/
% {\vphantom {{{r_n}} {{r_{\min }}}}} \right.
% \kern-\nulldelimiterspace} {{r_{\min }}}}} \right)^{ - v}}$.
% The resolution of ADCs is over $b = 1 \sim 3$ bits, and the values of $\rho$ is listed in Table I for $b \le 5$ bits.
This section provides numerical results to verify \emph{Lemma 1} in Sec. \ref{Section 3}. 
Consider a hexagonal cell with a radius of 1000 meters. 
%The users are assumed to be uniformly distributed over the cell, while the smallest distance between the users and the BS is ${r_{\min }} = 100$ meters.
User locations are generated independently and randomly in the cell by following uniform distribution, but the distance
between each user and the BS is at least 100 meters.
The path loss is expressed as $r_n^{ - v}$, where ${r_n}$ is the distance from the $n{\bf{th}}$ user to the BS and $v = 3.8$ is the path loss exponent.
Shadowing is modeled as a log-normal random variable ${s_n}$
with a standard deviation of 8dB. 
Thus the large-scale fading can be 
expressed as ${\beta _n} = {s_n}{\left( {{{{r_n}} \mathord{\left/
 {\vphantom {{{r_n}} {{r_{\min }}}}} \right.
 \kern-\nulldelimiterspace} {{r_{\min }}}}} \right)^{ - v}}$.
 The resolution of ADCs is 
 $b = 1 \sim 3$ bits, and the values of $\rho$ are listed in Table I for $b \le 5$ bits.
 \begin{table}
 \caption{$\rho$ for different resolution bits $b$} 
 \centering 
 \begin{tabular}{c c c c c c} 
 \hline 
 $b$   &1  &2  &3  &4  &5
 \\ \hline 
 $\rho$  &0.3634  &0.1175  &0.03454  &0.009497  &0.002499
 \\
 \hline
 \end{tabular}
 \end{table}
 
\begin{figure}[!h]
\centering
\includegraphics[width=3.5in, height=2.5in]{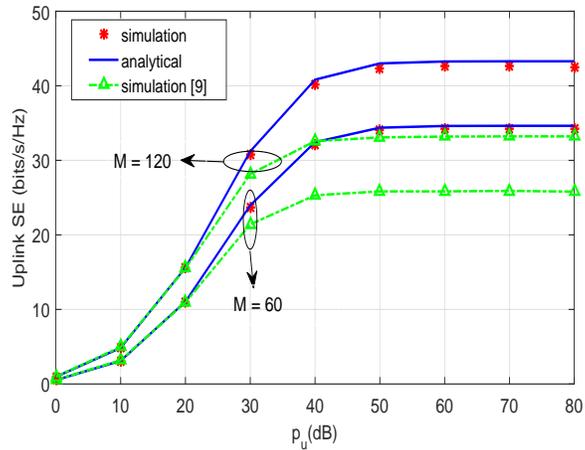}
\caption{Simulated and analytical SEs of massive MIMO with non-uniform resolution ADCs versus the transmit user power ${p_u}$.}
\label{fig1}
\end{figure}

 In Fig. \ref{fig1}, the simulated uplink SE along with the derived asymptotic equivalent in \emph{Lemma 1} are plotted against the user power ${p_u}$. The number of antennas $M$ is set as 60 and 120.  The number of users $K$ is 8. The resolution bits of each ADC is selected with equal probability over $b = 1\sim3$ bits. The simulated uplink SE from [9] is included for comparison. It is observed that the derived asymptotic equivalent of the uplink SE matches very well with the simulation results for the two different numbers of antennas $M$ and for all cases of the user power ${p_u}$. As the user power ${p_u}$ increases, our proposed MMSE receiver shows a much better performance than that proposed in \cite{Dong}. This is because, when the user power ${p_u}$ is high, the quantization noise becomes dominant than AWGN; our proposed MMSE receiver can suppress the quantization noise and AWGN at the same time, while the proposed MMSE receiver in \cite{Dong} can just suppress AWGN. It is also observed that as the user power ${p_u}$ increases, the uplink SE is saturated, because the power of quantization noise scales up with the user power ${p_u}$, as shown in Eq. (\ref{EQ4}).
\begin{figure}[!h]
\centering
\includegraphics[width=3.5in, height=2.5in]{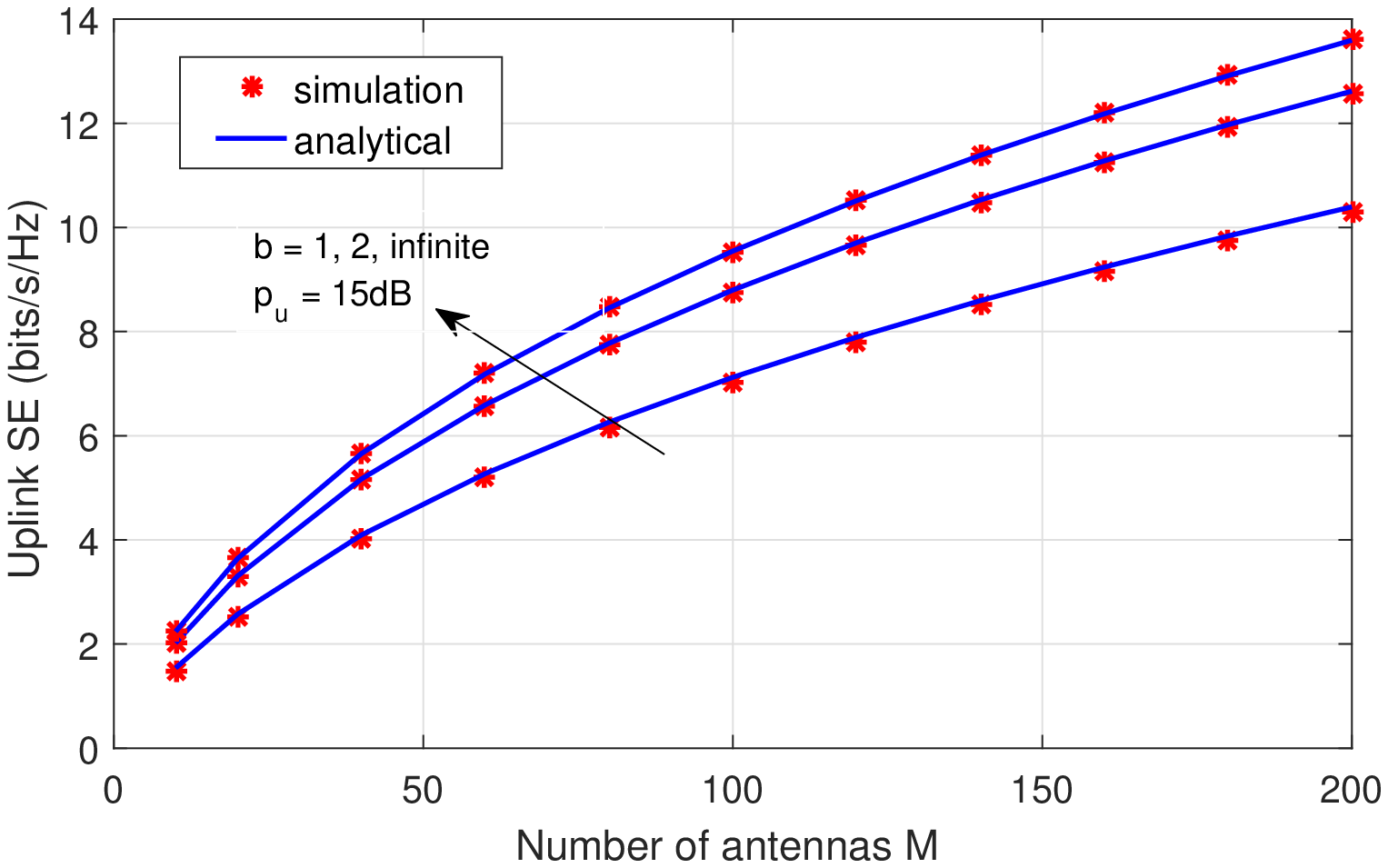}
\hfill
\includegraphics[width=3.5in, height=2.5in]{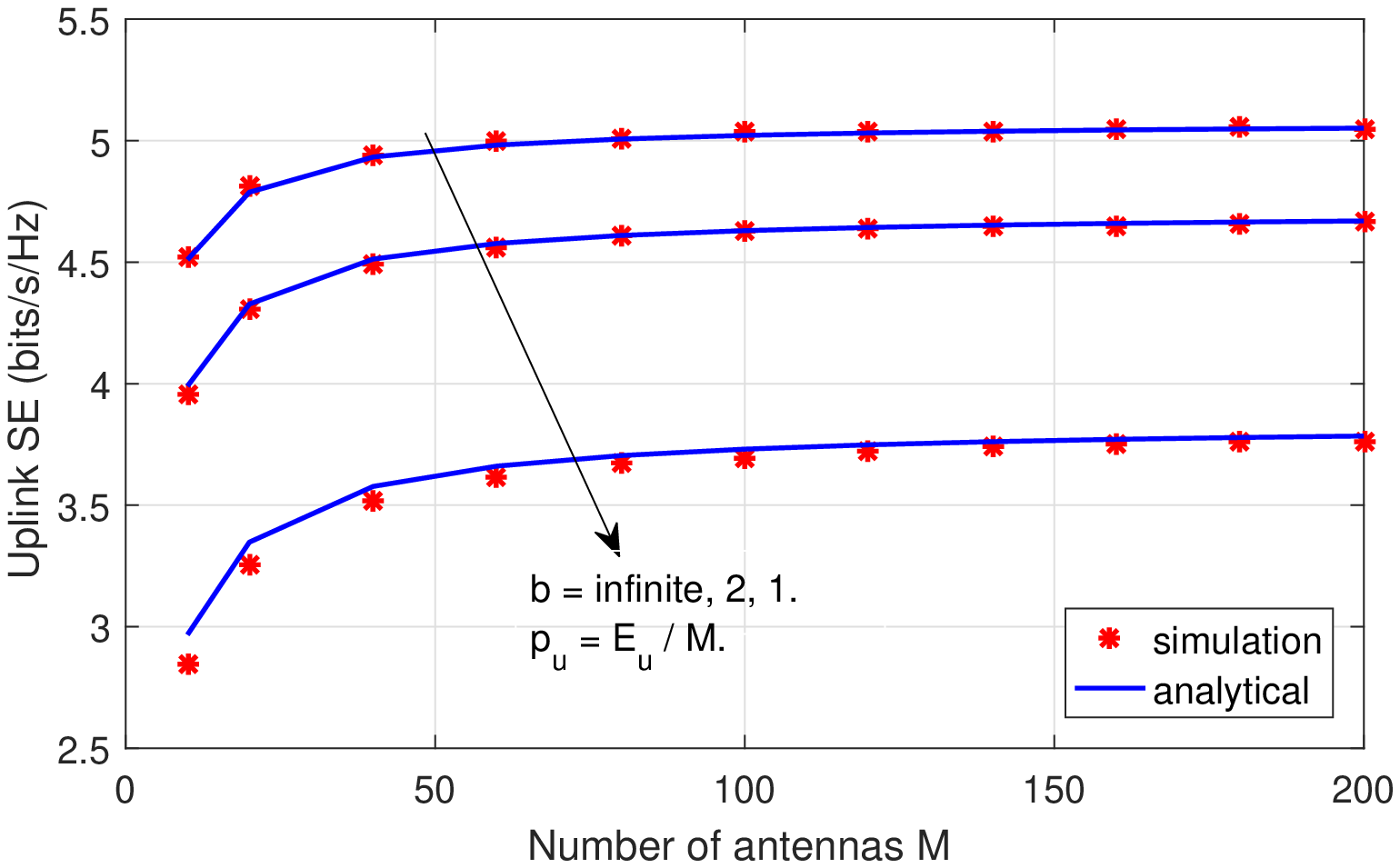}
\caption{Simulated and analytical SEs of massive MIMO with uniform-resolution ADCs versus the number of antennas $M$.}
\label{fig2}
\end{figure}

Fig. \ref{fig2} shows the uplink SE 
 versus the number of antennas with uniform-resolution ADCs as $1, 2,$ infinite bits. The number of users is 8. We can see that, in all cases of the number of antennas, 
 the simulation results 
 match very well the analytical results.
 With only $2$  bits, the SE nearly approaches 
 that with ideal ADCs
 (infinite resolution). For fixed values of ${p_u}$, all curves grow without bound 
 as $M$ increases, while for variable values of ${p_u} = {{{E_u}} \mathord{\left/
 {\vphantom {{{E_u}} M}} \right.
 \kern-\nulldelimiterspace} M}$, the corresponding curves eventually saturate  
 as $M$ increases. Note that ${E_u}$ is set as 30dB.
 %Fig. \ref{fig2} shows the uplink SE against the number of antennas with uniform-resolution ADCs as $1, 2,$ and infinite bits. The number of users is 8. We can see that, in all cases of the number of antennas $M$, a precise agreement between the simulation results and our analytical results in \emph{Lemma 1} can be found. 
 %And a higher resolution ADCs improve the SE. 
 %Moreover, with only $2$  bits, the SE nearly approaches the one of ideal ADCs with infinite resolution. For fixed values of ${p_u}$, all curves grow without bound by increasing $M$, while for variable values of ${p_u} = {{{E_u}} \mathord{\left/
 %{\vphantom {{{E_u}} M}} \right.
 %\kern-\nulldelimiterspace} M}$, the corresponding curves eventually saturate with the increasing of $M$. Note that ${E_u}$ is set as 30dB.
 
 \section{Conclusion}
 In this paper, AQNM is used to model the effect of the low-resolution ADCs. Based on the AQNM, we propose a high-performance MMSE receiver for massive MIMO systems with low-resolution ADCs, which can suppress both  AWGN and quantization noise at the same time. Using the proposed MMSE receiver and random matrix theory, we derive the asymptotic equivalent of the uplink SE. Numerical results show that the asymptotic equivalent is very tight. Moreover, the loss of the uplink SE brought by the low-resolution ADCs can be compensated by increasing the number of antennas at BS.

\appendices \label{Appendix}
\section{}
{\emph{Theorem 1 ([10])}}: Let $ {\bf{U}}  \in {{\mathbb{C}}^{M \times M}}$ and $ {\bf{Z}}  \in {{\mathbb{C}}^{M \times M}}$
be Hermitian nonnegative definite and $ {\bf{H}}  \in {{\mathbb{C}}^{M \times K}}$
be random with independent columns ${{\bf{h}}_k} \sim {\cal{C}}{\cal{N}} \left( {0,\frac{1}{M}{{\bf{R}}_k}} \right)$. Assume that ${\bf{U}}$ and ${{\bf{R}}_k}\left( {k = 1,2, \ldots ,K} \right)$, have uniformly bounded spectral norms (with respect to $M$). Then when $M,K \to \infty $ with fixed ratio and any $\rho  > 0$, 
\begin{equation}
\frac{1}{M}{\bf{tr}}\left( {{\bf{U}}{{\left( {{\bf{H}}{{\bf{H}}^H} + {\bf{Z}} + \rho {{\bf{I}}_M}} \right)}^{ - 1}}} \right) - \frac{1}{M}{\bf{tr}}\left( {{\bf{US}}\left( \rho  \right)} \right)
\overset{a.s.}{\longrightarrow}{0}, \nonumber
\end{equation}
where ${\bf{S}}\left( \rho  \right) = {\left( {\frac{1}{M}\sum\nolimits_{k = 1}^K {\frac{{{{\bf{R}}_k}}}{{1 + {\delta _k}\left( \rho  \right)}}}  + {\bf{Z}} + \rho {{\bf{I}}_M}} \right)^{ - 1}}$ and ${\delta _k}\left( \rho  \right)$ is defined as ${\delta _k}\left( \rho  \right) = {\lim _{t \to \infty }}\delta _k^t\left( \rho  \right)$, $k = 1, \ldots ,K$, where
\begin{equation}
\delta _k^{\left( t \right)}\left( \rho  \right) = \frac{1}{M}{\bf{tr}}\left( {{{\bf{R}}_k}{{\left( {\frac{1}{M}\sum\limits_{i = 1}^K {\frac{{{{\bf{R}}_i}}}{{1 + \delta _i^{\left( {t - 1} \right)}\left( \rho  \right)}}}  + {\bf{Z}} + \rho {{\bf{I}}_M}} \right)}^{ - 1}}} \right), \nonumber
\end{equation}
for $t=1,2, \ldots ,$ with initial value $\delta _k^{\left( 0 \right)} = {1 \mathord{\left/
 {\vphantom {1 \rho }} \right.
 \kern-\nulldelimiterspace} \rho }$ for all $k$.
 
 {\emph{Theorem 2 ([10])}}: Let $ {\Theta}  \in {{\mathbb{C}}^{M \times M}}$ be Hermitian nonnegative definite with uniformly bounded spectral norm (with respect to $M$). Under the same conditions as in {\emph{Theorem 1}},
\begin{equation}
\frac{1}{M}{\bf{tr}}\left( {{\bf{U}}{{\bf{A}}^{ - 1}}\Theta {{\bf{A}}^{ - 1}}} \right) - \frac{1}{M}{\bf{tr}}\left( {{\bf{US'}}\left( \rho  \right)} \right)
\overset{a.s.}{\longrightarrow}{0}, \nonumber
\end{equation}
where ${\bf{A}} = {\bf{H}}{{\bf{H}}^H} + {\bf{Z}} + \rho {{\bf{I}}_M}$ and ${\bf{S'}}\left( \rho  \right) \in {{\mathbb{C}}^{M \times M}}$ is 
\begin{equation}
{\bf{S'}}\left( \rho  \right) = {\bf{S}}\left( \rho  \right)\Theta {\bf{S}}\left( \rho  \right) + {\bf{S}}\left( \rho  \right)\frac{1}{M}\sum\limits_{k = 1}^K {\frac{{{{\bf{R}}_k}{{\delta '}_k}\left( \rho  \right)}}{{{{\left( {1 + {\delta _k}\left( \rho  \right)} \right)}^2}}}{\bf{S}}\left( \rho  \right)} . \nonumber
\end{equation}
${\bf{S}}\left( \rho  \right)$ and ${\delta _k}\left( \rho  \right)$ are defined in {\emph{Theorem 1}}. And ${\bf{\delta '}}\left( \rho  \right) = {\left[ {{{\delta '}_1}\left( \rho  \right), \ldots ,{{\delta '}_K}\left( \rho  \right)} \right]^T}$ is calculated as ${\bf{\delta '}}\left( \rho  \right) = {\left( {{{\bf{I}}_M} - {\bf{J}}\left( \rho  \right)} \right)^{ - 1}}{\bf{v}}\left( \rho  \right)$, where
\begin{equation}
\begin{array}{l}
{\left[ {{\bf{J}}\left( \rho  \right)} \right]_{kl}} = \frac{{\frac{1}{M}{\bf{tr}}\left( {{{\bf{R}}_k}{\bf{S}}\left( \rho  \right){{\bf{R}}_l}{\bf{S}}\left( \rho  \right)} \right)}}{{M{{\left( {1 + {\delta _l}\left( \rho  \right)} \right)}^2}}},{\rm{ 1}} \le k,l \le K\\
{\left[ {{\bf{v}}\left( \rho  \right)} \right]_k} = \frac{1}{M}{\bf{tr}}\left( {{{\bf{R}}_k}{\bf{S}}\left( \rho  \right)\Theta {\bf{S}}\left( \rho  \right)} \right),1 \le k \le K.
\end{array} \nonumber
\end{equation}
%\section{}

%\section*{Acknowledgment}


\ifCLASSOPTIONcaptionsoff
  \newpage
\fi

\end{document}